\documentclass[preprintnumbers, floatfix, letterpaper, superscriptaddress,nofootinbib, twocolumn]{revtex4}
\usepackage[dvipdfmx]{graphicx}
\usepackage{microtype}
\usepackage{amsmath}
\usepackage{amssymb}
\usepackage{subfigure}
\usepackage{hyperref}
\usepackage{url}
\usepackage{xcolor}
\usepackage{color}
\usepackage{mathrsfs}
\usepackage{amsfonts}
\usepackage{latexsym}
\usepackage{ragged2e}
\usepackage{epsfig}
\usepackage{textcomp}
\usepackage{phaistos}

\makeatletter
\renewcommand\@makefnmark{\hbox{\@textsuperscript{\normalfont\color{purple}\@thefnmark}}}
\renewcommand\@makefntext[1]{%
  \parindent 1em\noindent
            \hb@xt@1.8em{%
                \hss\@textsuperscript{\normalfont\@thefnmark}}#1}
\makeatother

\usepackage{caption}
\DeclareCaptionJustification{justified}{\leftskip=0pt \rightskip=0pt \parfillskip=0pt plus 1fil}
\def\beq{\begin{equation}}
\def\eeq{\end{equation}}

\def\mathbb{\Bbb}

\newcommand{\vp}{\varphi}

\newcommand{\pa}{\partial}
\newcommand{\df}{\dfrac}
\newcommand{\D}{\Delta }

\definecolor{vividviolet}{rgb}{0.62, 0.0, 1.0}
\definecolor{amaranth}{rgb}{0.9, 0.17, 0.31}
\definecolor{palatinateblue}{rgb}{0.15, 0.23, 0.89}
\definecolor{brightpink}{rgb}{1.0, 0.0, 0.5}
\definecolor{cornflowerblue}{rgb}{0.39, 0.58, 0.93}
\definecolor{deepcarminepink}{rgb}{0.94, 0.19, 0.22}
\definecolor{radicalred}{rgb}{1.0, 0.21, 0.37}
\colorlet{RED}{red}

\hypersetup{ linktoc=all,
    colorlinks, linkcolor={palatinateblue},
    citecolor={brightpink}, urlcolor={amaranth}
}
\newcommand{\changeurlcolor}[1]{\hypersetup{urlcolor=#1}}
\graphicspath{{Images/}}

\renewcommand{\d}[1]{\ensuremath{\operatorname{d}\!{#1}}}

\graphicspath{{Images/}}
\renewcommand{\d}[1]{\ensuremath{\operatorname{d}\!{#1}}}
\makeatletter
\def\@fnsymbol#1{\ensuremath{\ifcase#1\or $\textleaf$ \or $\PHplaneTree$
\else\@ctrerr\fi}}%

\makeatother

\def\sideremark#1{\ifvmode\leavevmode\fi\vadjust{\vbox to0pt{\vss
 \hbox to 0pt{\hskip\hsize\hskip1em
 \vbox{\hsize1.5cm\tiny\raggedright\pretolerance10000
 \noindent #1\hfill}\hss}\vbox to8pt{\vfil}\vss}}}%
\begin{document}

\title{Null Hypersurface Caustics, Closed Null Curves, and Super-Entropy}

\author{Sousuke \surname{Noda}}
\email{sousuke.noda@yukawa.kyoto-u.ac.jp}
\affiliation{Center for Gravitation and Cosmology, College of Physical Science and Technology, Yangzhou University, \\180 Siwangting Road, Yangzhou City, Jiangsu Province  225002, China}
 \affiliation{Division of Liberal Arts, Kogakuin University,
2665-1 Nakano-machi, Hachioji, Tokyo, 192-0015, Japan}

\author{Yen Chin \surname{Ong}}
\email{ycong@yzu.edu.cn}
\affiliation{Center for Gravitation and Cosmology, College of Physical Science and Technology, Yangzhou University, \\180 Siwangting Road, Yangzhou City, Jiangsu Province  225002, China}
\affiliation{School of Aeronautics and Astronautics, Shanghai Jiao Tong University, Shanghai 200240, China}

\begin{abstract}
Recently it was discovered that null hypersurfaces can develop caustics outside the event horizon of super-entropic Kerr-AdS black holes, in contrast to the usual Kerr-AdS case. 
In this work we explore a few more examples of black hole spacetimes in which such exterior caustics can develop. If a closed null curve is present, e.g., in the case of Taub-NUT and the ``transunital'' Kerr-AdS spacetimes, then it coincides with a null hypersurface caustic (NHC) of a minimal separation parameter. Thus a spacetime on the verge of forming closed timelike curves could develop a caustic.
Known examples of super-entropic black holes also have exterior NHC, although such spacetimes are free of closed null/timelike curves. 
Nevertheless the relationship between closed causal curves, NHC, and super-entropy is not straightforward. This is best illustrated with the BTZ black string, which for some choices of the warp factor in the extra dimension and the value of the charge, can be super-entropic. However, even those that are not super-entropic can admit NHC outside the horizon.
\begin{center}

\end{center}
\end{abstract}

\maketitle

\section{Introduction: Null Hypersurface Caustics}\label{1}

The Lorentzian signature of spacetime manifolds gives rise to the notion of causality, which is of great importance in general relativity. From the concept of black holes to the horizon problem in cosmology, the concept of causality plays a central role. More importantly, in any given physical system, we are often interested in predicting -- using physical laws in the form of evolution equations -- how the system evolves given initial and/or boundary conditions. Doing so requires that the Cauchy problem be well-posed, i.e., that the evolution is unique. Although we often think of slicing spacetime into a foliation of spacelike hypersurfaces \`{a} la Arnowitt-Deser-Misner (ADM) decomposition \cite{ADM1,ADM2} when discussing the Cauchy problem, for some purposes (notably in the context of black holes) it is more convenient to consider a three-dimensional null foliation \cite{1908.08739} or double-null foliation \cite{9510040, 1103.3538} instead.

The causal structure of a given spacetime $M$ is often said to be determined by the behavior of light cones. However, since the light cones at any point $p \in M$ live in the tangent space $T_p M$ instead of the spacetime manifold itself, sometimes we are really interested in how null curves in $M$ behave. The study of null hypersurfaces is therefore also a natural generalization of this. If the light cones ``fold up'' too much, caustics can develop in the null hypersurface. When this happens, the foliation is no longer good for practical purposes such as initial value problem in  numerical relativity. Worse still, sometimes it could mean that the causal structure of spacetime has some pathologies, such as a closed null curve. Surprisingly the exterior geometry of both asymptotically flat \cite{9803080} and asymptotically (anti-)de Sitter Kerr \cite{1909.06419} (hereinafter, ``Kerr-(A)dS'') black holes are free of such null hypersurface caustics (NHC), despite the complicated ways light rays can behave around rotating black holes. This remains the case even if the black holes are electrically charged. 

Imseis et al. \cite{2007.04354} recently showed that the so-called ``ultra-spinning'' super-entropic Kerr-AdS black hole, which is constructed by a nontrivial procedure when considering the limit $a \to L$ (see below), admits NHC outside the horizon. This prompts a question: could NHC be related to super-entropy in some way? On the other hand, if light cones open up too much there is a risk of admitting a closed timelike curve (CTC) in the spacetime, which would violate causality (whether this is a problem or not remains controversial). Therefore it is also interesting to investigate whether NHC is related to CTC. 

In Sec. (\ref{2}), we will investigate spacetimes with CTC and show that the boundary of the CTC region -- a closed null curve -- coincides with a NHC with a minimal separation parameter. Of particular interest is the ``transunital'' Kerr-AdS black hole, with $a > L$. 
In a sense, this illustrates how null hypersurface foliation is related to causality -- if caustic develops it could be a hint that there is a CTC, and thus (global) causality is violated. NHC can occur when closed causal curve\footnote{A closed curve is causal if it is either timelike or null.} is absent, so they are not equivalent notions.
In Sec. (\ref{3}), we will look at some examples of super-entropic black holes. The (limited) evidence suggests that all black holes which (for some parameters) can be super-entropic have NHC outside their horizon, but the converse does not hold. We end with some discussions in Sec. (\ref{4}).

\section{Null Hypersurface Caustics and Closed Null Curves}\label{2}

In this section we will investigate the connection between null hypersurface caustics and CTC, or more precisely the closed null curve that is a boundary of spacetime regions with CTC. It is instructive to begin with the ``transunital'' black holes, which are just Kerr-AdS black hole with angular momentum parameter $a$ larger than the asymptotic AdS curvature length scale $L$, since the mathematics is rather similiar with the ``cisunital'' (the usual $a<L$) case and the super-entropic case ($a=L$, but with nontrivial topological construction). The notion of NHC will be defined as we work through this example. {(The readers should refer to \cite{2007.04354} for more detailed explanations.)} 

\subsection{Transunital Kerr-AdS Black Hole}

The metric of the Kerr-AdS$_5$ spacetime with a single rotation axes in 5 dimensions\footnote{The reason this was explored in 5 dimensions is so that it can be applied in holography \cite{1906.01169, 1911.08222}. A practical reason is because the horizon function $\Delta_r$ vastly simplifies in 5 dimensions, namely the second term $-2M$ has no $r$-dependence, while it is $-2Mr$ in 4 dimensions.} is \cite{1906.01169, 1911.08222,2005.03869,2006.09385}:
\begin{flalign}
g[\text{KAdS}(a>L)]=&-\df{\Delta_r}{\rho^2}\left[\d t-\df{a}{\Xi} \sin^2{\theta} \d\varphi \right]^2 \\ \notag &+\df{\sin^2{\theta} \Delta_\theta}{\rho^2}\left[a \d t -\df{r^2+a^2}{\Xi}\d\varphi \right]^2 
\\ \notag &+\df{\rho^2}{\Delta_r}\d r^2+\df{\rho^2}{\Delta_\theta}  \d\theta^2 +r^2\cos^2\theta \d\psi^2,
\end{flalign}
where 
\begin{align}
\notag &\rho^2:=r^2+a^2\cos^2{\theta},\\
\notag &\Delta_r:=(r^2+a^2)\left(1+\df{r^2}{L^2}\right)-2M,\\
\notag &\Delta_\theta:=1-\df{a^2}{L^2}\cos^2{\theta},\\
 &\Xi:=1-\frac{a^2}{L^2}.
\end{align}
Here the angular coordinates are Hopf coordinates on the topological 3-sphere (with $\varphi, \psi \in (0,2\pi)$ and $\theta \in (0, \pi/2)$).
Note that $M$ and $a$ are, respectively, the mass and spin parameters of the black hole. The conserved \emph{physical} mass and angular momentum that enter the thermodynamical laws of this black hole are \cite{0408217}
\beq
{\cal{M}}=\df{\pi M (2+\Xi)}{4 \ell_\text{B}^3 \Xi^2},\ \ {\cal{J}}=\df{\pi M a}{2\ell_\text{B}^3 \Xi^2},
\eeq
where $\ell_\text{B}$ is the gravitational length scale in the bulk. The physical angular momentum  
to mass ratio ${\cal{A}}$ is defined as ${\cal{J}}={\cal{A}} {\cal{M}}$, hence, 
\beq
{\cal{A}}=\df{2a}{2+\Xi}=\df{2a}{3-(a/L)^2}.
\eeq

Following \cite{2005.03869}, we refer to those black holes with $a>L$ as ``transunital'', while those with $a<L$ as ``cisunital''. There exists a mapping \cite{0601002,0604125} between the metrics of transunital and cisunital black holes, such that the geometries are \emph{locally} equivalent. However, this is not a global equivalence (e.g., the full ranges of angular coordinates are not preserved under such a coordinate transformation) and so it does not preserve global quantities. It is thus not surprising that transunital black holes can admit closed causal curves but their cisunital cousins do not.

Note that ${\cal{A}}/L$ is a monotonic increasing function of $a/L$ and for $a/L=1$, ${\cal{A}}/L$ is also unity. Of course the metric is not defined exactly at $a=L$, but as we shall see, cosmic censorship does not allow $a$ to come close to $L$ anyway (from either side); see \cite{1906.01169} for details. In fact, the $a=L$ case can only be made sense of by nontrivial topological identification, which gives rise to the super-entropic case that we will discuss in Sec. (\ref{3}).

Returning to our case, we note that for $a/L \rightarrow \sqrt{3}$ $(\Xi \rightarrow -2)$, ${\cal{A}}/L \rightarrow+\infty$.
Depending on the spin parameter $a$, this spacetime may not have an event horizon. 
To obtain the condition for the horizon to exist (that is, for cosmic censorship condition to hold), we only need to examine 
the function $\Delta_r$, which is a quadratic in $r^2$:
\beq
\df{\D_r}{L^2}=\bar{r}^4+\left(1+\df{a^2}{L^2}\right) \bar{r}^2 +\df{a^2-2M}{L^2},
\eeq
where we have introduced the re-scaled radial coordinate $\bar{r}:=r/L$.
For this equation to have a positive real solution, $a$ needs to satisfy the bound $a^2<2M$.
In terms of $\Xi$, this condition can be written as
\beq
(1-\Xi)<\frac{2M}{L^2} =\df{8\ell_\text{B}^3 {\cal{M}}}{\pi L^2} \df{\Xi^2}{2+\Xi} =: \mu  \df{\Xi^2}{2+\Xi},
\eeq
where $\mu$ is a kind of dimensionless mass parameter. Then, the cosmic censorship condition yields
\beq
(\mu+1)\Xi^2 +\Xi -2 >0.
\eeq
This inequality gives 
\beq \label{ccc}
\Xi<\Xi^{(-)},\ \ \Xi^{(+)}<\Xi,
\eeq
where 
\beq
\Xi^{(\pm)}=\df{-1\pm \sqrt{1+8(\mu+1)}}{2(\mu+1)}.
\eeq
Note that $\Xi^{(-)}<0$ and $\Xi^{(+)}>0$.

For some of the subsequent calculations, we will need the inverse metric. To this aim we introduce the following quantity:
\beq
D:=g_{t\vp}^2 -g_{tt} g_{\vp \vp}=\D_r \D_\theta \df{\sin^2{\theta}}{\Xi^2}.
\eeq
Using this, the component of the inverse metric tensor we need can be obtained as
\beq
g^{tt}=-\df{g_{\vp\vp}}{D}=\df{1}{\rho^2}\left[\df{a^2  \sin^2{\theta}}{\D_\theta}-\df{ (r^2+a^2)^2}{\D_r}\right],
\eeq
the remaining components that we will use are $g^{rr}$ and $g^{\theta\theta}$, which are just the inverse of the components of the metric tensor.

Let us now study the null hypersurfaces of this spacetime to investigate its caustics structure by following the method of \cite{9803080, 1909.06419, 2007.04354}.
We start by introducing the ingoing and outgoing Eddington-Finkelstein coordinates; these are defined in terms of a ``generalized tortoise coordinate'' $r_*(r,\theta)$ with angle dependence: 
\beq
v=t+r_*(r,\theta),\ \ u=t-r_*(r,\theta).
\label{eq:vu}
\eeq
The exact form of $r_* (r,\theta)$ will be determined by Eq.~\eqref{eq:PDE}.  
In terms of these coordinates, the null hypersurfaces are given by
\beq
v=\text{const},\ \ u=\text{const}.
\label{eq:vuconst}
\eeq
Therefore, the null hypersurface defined by $v=\text{const}$ satisfies the equation
\beq
g^{\mu\nu}\pa_\mu v \pa_\nu v =g^{tt}+g^{rr}(\pa_r r_{*})^2+g^{\theta \theta}(\pa_\theta r_{*})^2=0.
\label{eq:PDE}
\eeq
Note that if we write the above partial differential equation (PDE) in terms of $u$, we obtain the same $r_{*}$.
Therefore, once the PDE \eqref{eq:PDE} is solved, we can substitute 
the solution $r_{*}=r_*(r,\theta)$ into Eq.~\eqref{eq:vuconst} to obtain the null hypersurfaces.
Inserting $g^{tt}$ into the PDE, we obtain the separable form 
\beq
\left[\df{a^2  \sin^2{\theta}}{\D_\theta}-\df{ (r^2+a^2)^2}{\D_r}\right]+\D_r (\pa_r r_*)^2+ \D_\theta (\pa_\theta r_{*})^2=0
\eeq
Introducing the so-called ``constant of separation'' ${a^2}\lambda$ (hereinafter, $\lambda$ is referred to as the separation constant), we obtain the following system of PDEs:
\beq
\pa_r r_{*}=\df{Q(r)}{\Delta_r},\ \ \pa_\theta r_{*}=\df{P(\theta)}{\D_\theta},
\eeq
where 
\beq
Q(r)^2:=(r^2+a^2)^2-a^2\lambda \D_r,\ \ P(\theta)^2:=a^2\left(\D_\theta \lambda -\sin^2{\theta}\right).
\label{eq:PQ}
\eeq

Writing down the exact differential $\d r_{*}=\pa_r r_{*} \d r + \pa_\theta r_{*} \d\theta$, 
we can write
\beq
\d r_{*}=\df{Q}{\D_r} \d r +\df{P}{\D_\theta} \d\theta.
\label{eq:dr*}
\eeq
If we integrate this, there will be an integration constant, which we will denote by $(a^2/2) g(\lambda)$ with an arbitrary function $g(\lambda)$:
\beq
r_*=\int  \df{Q}{\D_r} \d r +\int \df{P}{\D_\theta} \d\theta +\df{a^2}{2}g(\lambda).
\label{eq:general}
\eeq
 To find a general solution $r_*(r,\theta)$, we assume $\lambda$ is also a variable: $r_*=\eta(r,\theta,\lambda)$, hence the exact differential is 
\beq
\d\eta=\df{Q}{\Delta_r} \d r + \df{P}{\Delta_\theta} \d\theta + \df{a^2}{2}F \d\lambda,
\label{eq:deta}
\eeq
where $\pa_\lambda \eta$ is written as $(a^2/2) F$ with 
\beq
 F(r,\theta,\lambda)=\int_r^{\infty} \df{1}{Q(r^{\prime},\lambda)} \d r^{\prime} +\int_0^{\theta} \df{1}{P(\theta^{\prime},\lambda)}\d\theta^{\prime} +g^{\prime}(\lambda).
 \eeq
If $\lambda=\text{const}$, Eq.~\eqref{eq:deta} reduces back to Eq.~\eqref{eq:dr*}. However, this result can be obtained by requiring $F=0$ even for $\lambda \neq \text{const}$. The condition $F=0$ determines the $(r,\theta)$-dependence of $\lambda$ for any choice of $g(\lambda)$ and uphold the original exact differential \eqref{eq:dr*}. Substituting $\lambda=\lambda(r,\theta)$ into Eq.~\eqref{eq:general}, 
the most general solution to Eq.~\eqref{eq:PDE} is obtained.
%

The condition $F=0$ implies $\d F=0$, {which gives}
 \beq
 0=(\pa_{\lambda}F) \d\lambda + (\pa_r F) \d r+ (\pa_\theta F)\d\theta,
 \eeq
 and we can write it as
 \beq
 \nu \d\lambda = -\df{\d r}{Q}+\df{\d\theta}{P},\ \ \text{with}\ \ \nu:=-\pa_\lambda F.
 \label{eq:nu}
 \eeq
 Making use of this and Eq. \eqref{eq:dr*}, we can rewrite the metric as
 \begin{flalign}
 \d s^2&=\df{\D_r \D_\theta}{\Xi^2 R^2}(\d r_{*}^2-\d t^2)+g_{\vp \vp}\d\varpi^2 +\df{\nu^2 P^2 Q^2}{\Xi^2 R^2} \d\lambda^2 \\ \notag &+r^2 \cos^2{\theta} \d\psi^2 \\ \notag
\\&=:\df{\D_r \D_\theta}{\Xi^2 R^2}(\d r_{*}^2-\d t^2)+\d h^2,
 \label{eq:metric}
 \end{flalign}
in which we have introduced $\d\varpi:=\d\vp -\omega \d t$, with $\omega:=-g_{t\vp}/g_{\vp\vp}$ and $R^2:=g_{\vp\vp}/\sin^2{\theta}$. The derivation of this form of the metric is provided in Appendix A.

Under the condition $\d r_{*}=\pm \d t$, the spacetime metric degenerates into:

 \beq
 \d h^2=g_{\vp \vp}\d\varpi^2 +\df{\nu^2 P^2 Q^2}{\Xi^2 R^2} \d\lambda^2 +r^2 \cos^2{\theta} \d\psi^2.
 \eeq
Since the determinant is the square of the volume element of this degenerate metric, 
the points where it becomes zero correspond to the caustics.
For the present metric, the condition for the caustics is thus given by
 \beq
  \df{r^2}{4}\sin^2({2\theta})\  \df{\nu^2 P^2 Q^2}{\Xi^2 }=0.
  \label{eq:caustics}
 \eeq
 
  We analyze Eq. \eqref{eq:caustics} for ingoing null hypersurface ($\lambda =\text{const}$ and decreasing $r$). Firstly, we note that from Eq. \eqref{eq:nu}, for fixed $\lambda$, for the case that decreasing $\theta$ 
 gives decreasing $r$, we have $P>0$. Therefore, $Q=0$ is (at least) a sufficient condition that gives rise to the caustics.

In the unit of $L$, we can define the following dimensionless quantities
\begin{widetext}
\begin{equation}
\bar{Q}^2(\bar{r};\Xi,\lambda,\mu):= Q^2/L^4 =(\bar{r}^2+1-\Xi)^2 -\lambda(1-\Xi)\left[(\bar{r}^2+1-\Xi)(1+\bar{r}^2)-\mu \df{\Xi^2}{2+\Xi}\right],
\end{equation}
\begin{equation}
\bar{P}^2(\theta;\Xi,\lambda):=P^2/L^2=(1-\Xi)\left[ \left(\sin^2{\theta}+\Xi \cos^2{\theta}\right) \lambda-(1-\Xi)\sin^2{\theta} \right].
\end{equation}
\end{widetext}

\begin{figure}[h!] 
  \centering
  \includegraphics[width=7.8cm]{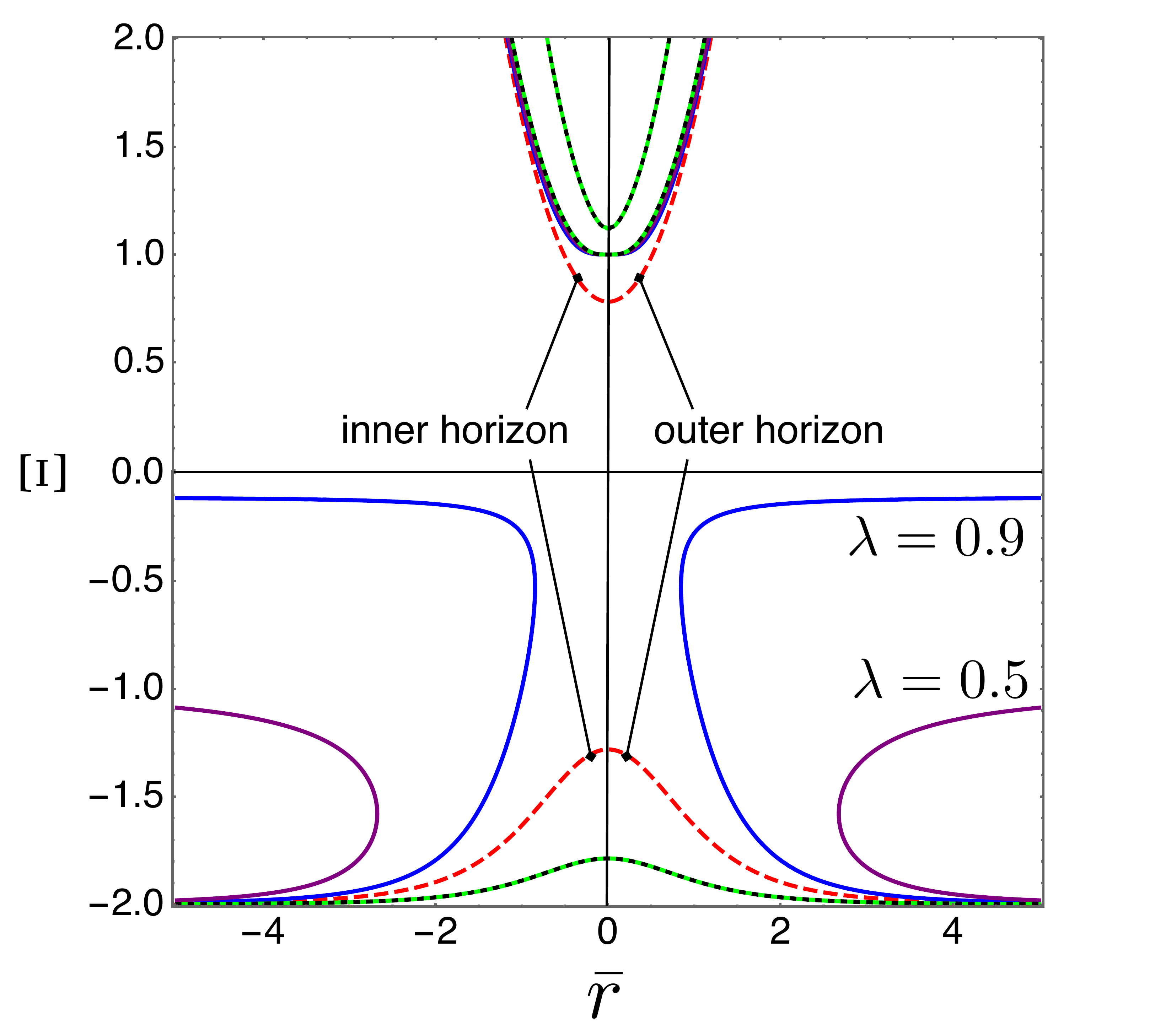}
  \includegraphics[width=7.8cm]{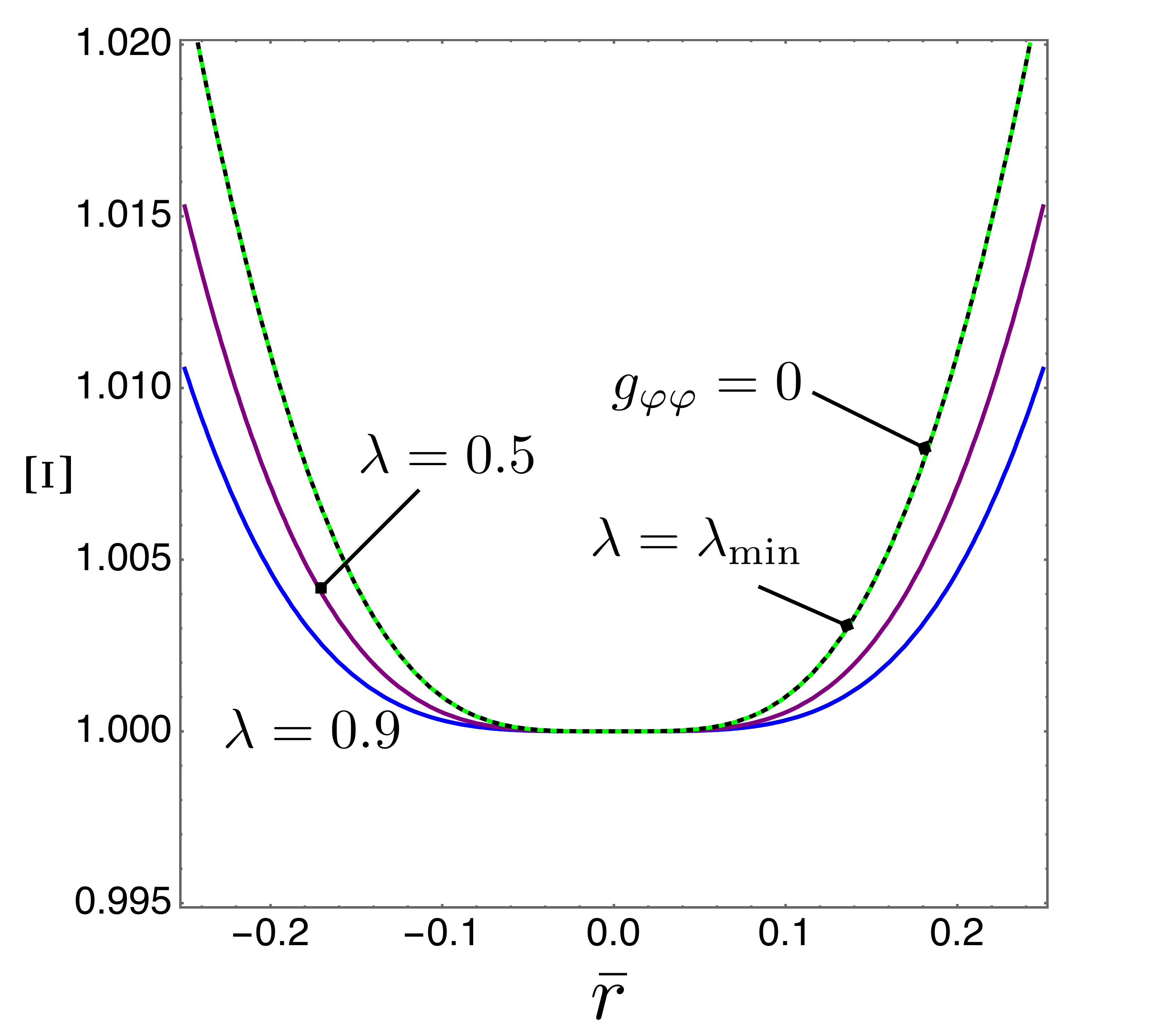}
  \caption{\footnotesize{Contour map of  ${Q}=0$ for $\lambda=0.9$ (blue), $\lambda=0.5$ (purple), $\lambda=\lambda_\text{min}$ (green), horizon (red dashed curve), and $g_{\vp \vp}=0$ 
  (black dotted) with $\theta=\pi/6$ shown in the $(\bar{r},\Xi)$-plane, where $\Xi:=1-a^2/L^2$. 
	Note that for a given value of $\theta$, there can be multiple caustics. For example the $\lambda=\lambda_\text{min}$ case here gives two caustic curves for the cisunital case, both of which coincide with a closed null curve. Both are located inside the event horizon.
 We also see that the caustics are outside the event horizon for transunital black holes ($\Xi<0$) 
 but not for cisunital black holes ($\Xi>0$). Note that there are some ranges of $\Xi$ in which there is no black hole. This is due to the censorship condition given in Eq. (\ref{ccc}). The lower panel is the zoom-in contours for the cisunital case.}}
  \label{fig:effective}
\end{figure}

We shall plot $\bar{Q}=0$ on the $(\bar{r},\Xi)$-plane for given values of $\lambda$ and $\mu$, which give us the NHC. 
We also plot the horizon location ($\D_r=0$) curve on the same plane. 
Moreover, we can now check for the existence of closed null curve and its position in the same plane. These are shown in Fig. (\ref{fig:effective}).

From the equation for $P$, namely Eq. \eqref{eq:PQ}, we get a lower bound of $\lambda$ for given $\theta$ and $\Xi$ as 
\beq
\lambda_\text{min}=\df{\sin^2{\theta}}{\Delta_\theta}.
\eeq
If we substitute this into $\bar{Q}^2$, the caustics condition $\bar{Q}^2=0$ requires that the expression
\beq
(\bar{r}^2+1-\Xi)^2\Delta_\theta -(1-\Xi)\sin^2{\theta}\left[ (\bar{r}^2+1-\Xi)(1+\bar{r}^2)-\mu\df{\Xi^2}{2+\Xi}\right]
\eeq
must vanish.
On the other hand, the closed null curve condition $g_{\vp \vp}=0$ can be written using $\Xi$ as
\begin{widetext}
\beq
\df{\sin^2{\theta}}{(\Xi^2/L^2)\left[\bar{r}^2+(1-\Xi)\cos^2{\theta}\right]}\left\{(\bar{r}^2+1-\Xi)^2\Delta_\theta -(1-\Xi)\sin^2{\theta}\left[ (\bar{r}^2+1-\Xi)(1+\bar{r}^2)-\mu\df{\Xi^2}{2+\Xi} \right] \right\}=0.
\eeq
\end{widetext}

Therefore, $g_{\vp \vp}=0$ and $\bar{Q}^2(\bar{r},\Xi,\lambda_\text{min},\mu)=0$ give the same curve.
That is to say, the boundary of the CTC region coincides with a null hypersurface caustic of a minimal separation parameter.

Before we move on to other examples, let us comment on the CTCs in the transunital black hole spacetime. As discussed in \cite{1906.01169}, the angular coordinate $\psi$ becomes timelike for $\theta < \theta_a := \text{arccos}(L/a)$ (These regions can be excised so that the black hole topology is that of a sphere with two punctures). Thus, there are CTCs for the region close to the poles. These are \emph{not} the CTCs we discussed here, which is in the $\vp$-direction. 
It can be checked that the metric coefficient
\begin{equation}
g_{\vp\vp} = \frac{\sin^2\theta}{\rho^2\Xi^2}\left[-\Delta_r a^2 \sin^2\theta + \Delta_\theta (r^2+a^2)^2\right]
\end{equation}
can become negative. This always happens at sufficiently large $r$. To see this, one can examine the leading term in $\Delta_r a^2 \sin^2\theta$ and $\Delta_\theta (r^2+a^2)^2$, that is, the coefficients of the $r^4$ terms. From this we can conclude that $g_{\phi\phi} <0$ for $a>L$ at large enough $r$. 

In our discussion above, the closed null curve $g_{\vp\vp} = 0$ is the one that corresponds to the caustics.
It is not clear whether such closed causal curves are pathological from holographic point of view, since the (conformal) boundary metric (see \cite{1906.01169}) is free of CTC in the $\vp$-direction. 
This is in contrast to the conical defect AdS$_3$ spacetime examined in \cite{1508.04440}, whose CTCs extend to the boundary. Yet even in that case the holographic dual seemingly admits a consistent and controllable evolution even without imposing additional consistency constraints.
In any case the physical relevance (or the lack thereof) is interesting but it is beyond the scope of work to discuss these issues in depth: our objective is merely to discuss the relationship between closed causal curve and NHC.

The aforementioned relationship between caustics and closed null curve also exist when they are \emph{inside} a black hole horizon. This can be checked, e.g., in the case of asymptotically flat Kerr black hole, in the region $r < 0$ ``below'' the ring singularity. For cisunital and transunital Kerr-AdS$_5$ black holes, see Fig. (\ref{fig:effective}).
It is worth noting at this point that the property that NHC can occur outside a black hole horizon is \emph{not} unique to super-entropic black holes, since transunital black holes are not super-entropic in the usual sense, that is, with respect to thermodynamic volume (see Sec.(\ref{3}) for definition). They can, however, be super-entropic with respect to mass \cite{2006.09385}. 

The method of searching for caustics explained thus far can be applied to various axisymmetric spacetimes. Therefore in the following parts of this work, we shall employ this method to discuss the caustics in a few other interesting spacetimes, with the hope to learn some shared properties and differences between them.

\subsection{Taub-NUT}

The Taub-NUT spacetime is a peculiar geometry that admits closed causal curves, but it also shares some similarities with Kerr black holes (in fact, it can be written in a Boyer-Lindquist-like coordinates in which the solution looks like a ``twisting'' black hole with the two hemispheres rotating in opposite directions \cite{1610.05757, 1609.09721}), so it provides another arena that we can explore to check the relationship between null hypersurface caustics and closed causal curves. 

The metric tensor of the Taub-NUT spacetime can be written in the 
coordinates $(\tau,r,\theta,\vp)$ as \cite{Bonner}
\begin{flalign}
g[\text{T-N}]&=-U(r)\left[\d\tau +4a \sin^2{(\theta/2)} \d\vp \right]^2+\df{\d r^2}{U(r)}\\ \notag &+(r^2+a^2)(\d\theta^2+\sin^2\theta \d\vp^2),
\end{flalign}
where 
\beq
U(r)=\df{r^2-2Mr-a^2}{r^2+a^2}.
\eeq
Here $a$ is \emph{not} a rotation parameter but the NUT charge, which has no Newtonian analog.

Alternatively, in $(t,r,\theta,\vp)$ coordinates, in which $t=\tau+2a \vp$, we can write the metric as
\begin{flalign}
g[\text{T-N}]=&-U\d t^2 -4aU\cos{\theta} \d t  \d\vp +\df{\d r^2}{U}+(r^2+a^2)\d\theta^2 \\ \notag &+\left[ -4a^2 U \cos^2{\theta}+(r^2+a^2)\sin^2{\theta}\right]\d\vp^2.
\end{flalign}
To evaluate the null hypersurface, we need the inverse metric component $g^{tt}$:
\begin{flalign}
g^{tt}&=-\df{g_{\vp\vp}}{g_{t\vp}^2-g_{tt}g_{\vp \vp}}=\df{4a^2 U \cos^2{\theta}-(r^2+a^2)\sin^2{\theta}}{(r^2+a^2)U\sin^2{\theta}}\\ \notag &=\df{4a^2}{(r^2+a^2)}\cot^2{\theta}-\df{1}{U(r)}.
\end{flalign}
For $v=t+r_*$, the null hypersurface is given by
\beq
g^{\mu\nu}\pa_\mu v \pa_{\nu} v= g^{tt}+g^{rr} (\pa_r r_*)^2+ g^{\theta\theta} (\pa_\theta r_*)^2=0,
\eeq
which yields
\begin{flalign}
(\pa_r r_*)^2&=\df{r^2+a^2-a^2\lambda U}{U^2 (r^2+a^2)} \\ \notag &=\df{(r^2+a^2)^2-a^2\lambda (r^2-2Mr -a^2)}{U^2(r^2+a^2)^2},\\
\end{flalign}
and
\begin{equation}
 (\pa_\theta r_*)^2=a^2(\lambda -4\cot^2{\theta}).
\end{equation}
From the second equation, we get the bound on the separation constant
\beq
\lambda \geqslant 4 \cot^2{\theta}:=\lambda_{\text{min}}.
\eeq

\begin{figure}[h!] 
  \centering
  \includegraphics[width=8cm]{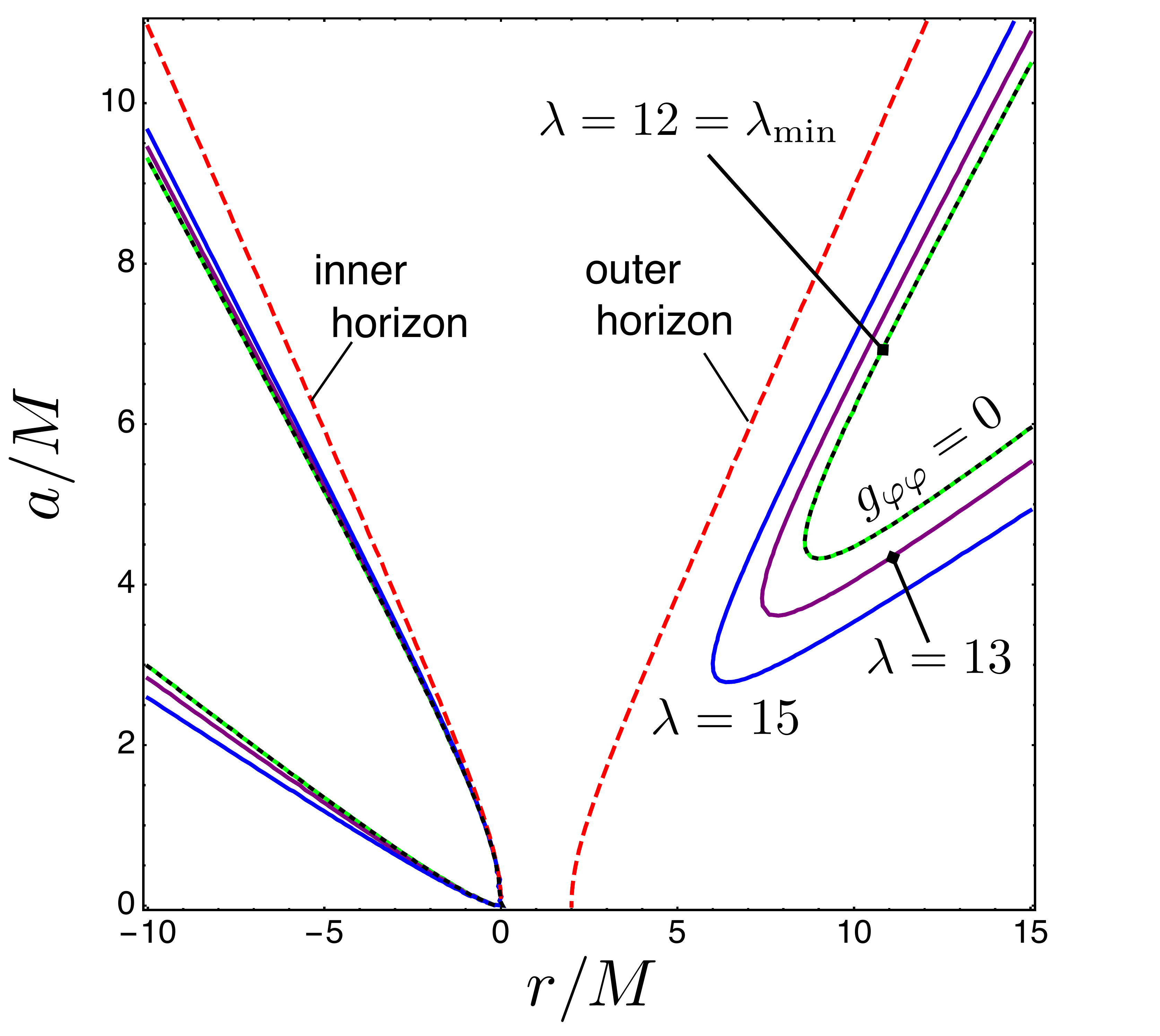}
  \caption{\footnotesize{Contour map of the caustics $[r^2+a^2-\lambda U(r)]=0$ for different separation constants $\lambda$ (blue, purple, green), horizon radii (dashed red), 
  and the boundary of the closed timelike curve region, $g_{\vp\vp}=0$ (dotted black), for the choice $\theta=\pi/6$, in the Taub-NUT spacetime.}}
  \label{fig:ra}
\end{figure}

In Fig. (\ref{fig:ra}), the caustics condition $(r^2+a^2)^2-a^2\lambda (r^2-2Mr -a^2)=0$, the horizon condition $r^2-2Mr-a^2=0$, as well the CTC condition (null closed curve $g_{\vp\vp}=0$) for given $\theta$ on $(r,a)$-plane are plotted. In these plots, for definiteness we have set $\theta=\pi/6$, hence, $\lambda_\text{min}=12$. 

Once again, we observe that the null closed curve coincides with the NHC with minimal separation constant. This is true both outside and inside the horizon. Note that the negative $r$ region is connected with the positive $r$ region by the 2-surface of finite area at $r=0$ (at which the curvature is finite and hence not a singularity) \cite{1610.06135}.

\subsection{Tipler cylinder}
As another example of spacetime with CTC, we consider one of the archetypal example of time machine: the Tipler cylinder \cite{Tipler}, 
which is an infinitely long and massive spinning object. The metric for the whole region of the spacetime is written in the following form: 
\beq
g[\text{TIP}]=-F \d t^2 + 2M \d t \d\vp +H(\d r^2+\d z^2) +L \d\vp^2.
\eeq
The $z$-axis corresponds to the spin axis of the cylinder. 
The inner region of the surface whose radius is $R$ is given by
\beq
F=1,\ M=ar^2,\ H=e^{-a^2 r^2},\ L=r^2(1-a^2 r^2),
\eeq
where $a$ is the angular velocity of the cylinder.
Separating the PDE $g^{\mu\nu}\pa_\mu v \pa_\nu v =0$ with the separation constant $\lambda$, 
we obtain 
\begin{align}
&(\pa_r r_*)^2=\df{1-a^2 r^2 -\lambda e^{a^2 r^2}}{e^{a^2 r^2}},\\
&(\pa_z r_*)^2=\lambda.
\end{align}
Considering these equations and $L$, we will see that caustics with $\lambda=\lambda_\text{min}=0$ 
gives the boundary of the CTC region $g_{\vp\vp}=0$.

\begin{figure}[h!] 
  \centering

  \includegraphics[width=7.85cm]{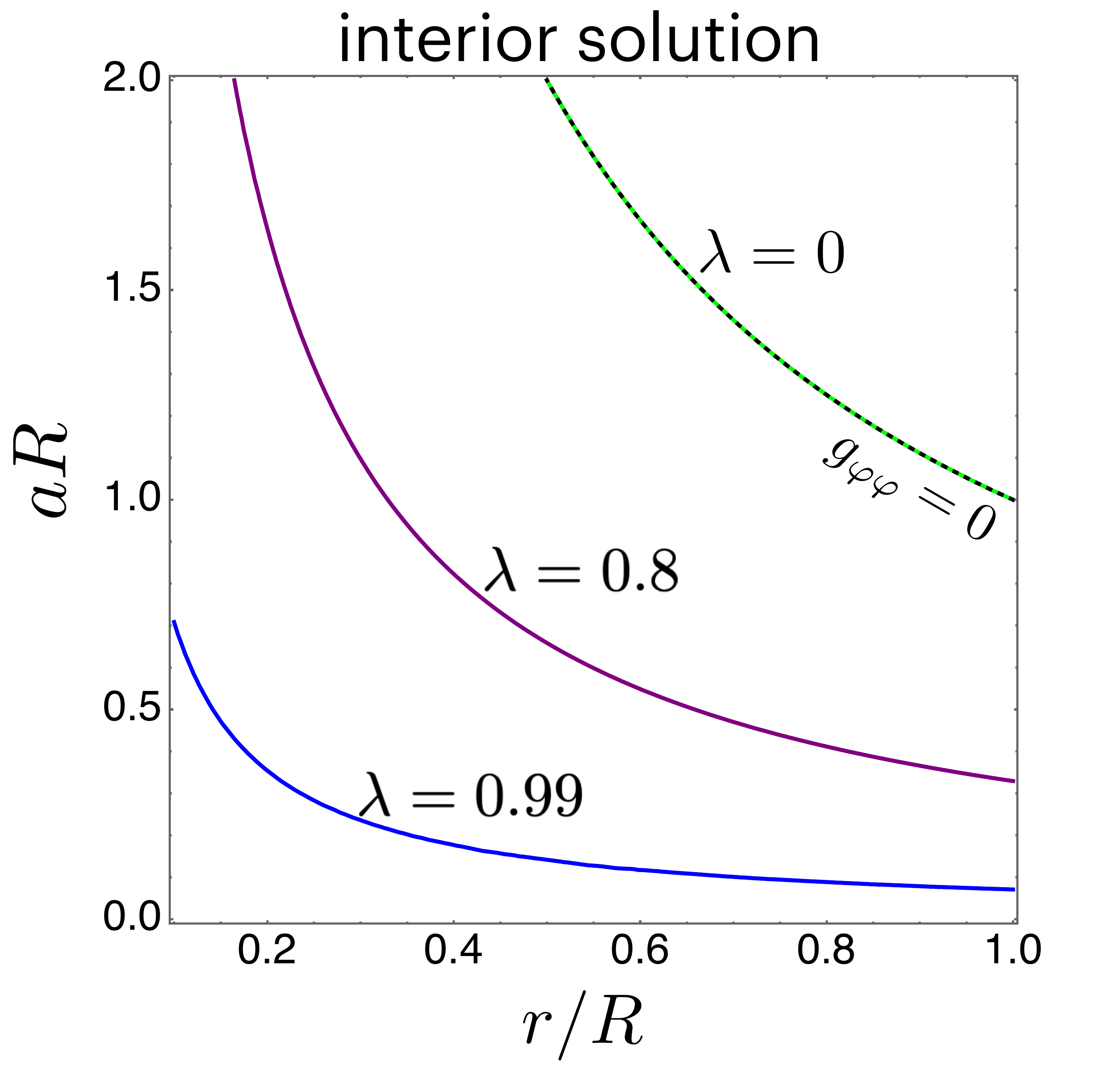}

\bigbreak

  \includegraphics[width=8.05cm]{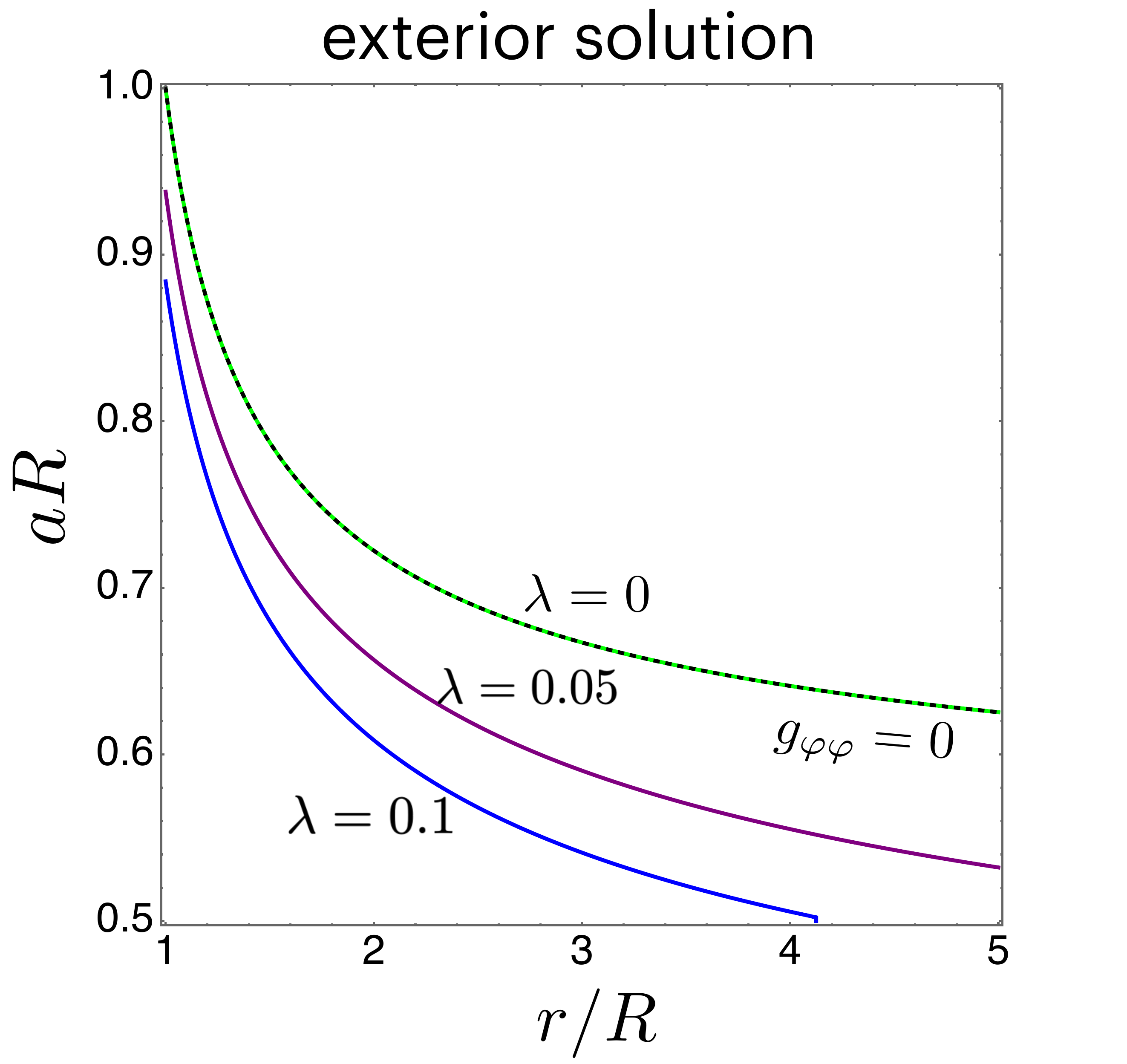}

  \caption{\footnotesize{{NHC curves and $g_{\vp\vp}=0$ (black dotted) for interior solution (upper panel) and exterior solution (lower panel) for several values of $\lambda$ in the Tipler cylinder spacetime.}}}
  \label{fig:Tipler}
\end{figure}

The exterior solution is classified into the following three classes depending on the radius $R$: $0<aR<1/2$, $aR=1/2$, and $1/2<aR < 1$. The upper bound $aR<1$ stems from the fact that the 
surface velocity of the cylinder is supposed to be slower than the speed of light. According to \cite{Tipler}, 
only the third case possesses CTC, therefore, we focus on this case here. The metric functions are
\begin{align}
&F=\df{r \sin{(\beta -\gamma)}}{R \sin{\beta}},\ M=\df{r\sin{(\beta +\gamma)}}{\sin{2\beta}},\\
&H=e^{-a^2 R^2}\left(\df{r}{R}\right)^{-2a^2 R^2},\ L=\df{Rr \sin{(3\beta+\gamma)}}{2\sin{2\beta}\cos{\beta}},
\end{align}
with
\beq
\beta=\tan^{-1}{\left(4a^2 R^2-1\right)^{1/2}}  ,\ \gamma=\left(4a^2 R^2-1\right)^{1/2}\log{(r/R)}.
\eeq
The PDE for the NHC yields
\begin{align}
&(\pa_r r_*)^2=\df{R\sin{(3\beta+\gamma)}-2\lambda e^{a^2 R^2}(r/R)^{2a^2R^2}r\sin{2\beta}\cos{\beta}}{2r \sin{2\beta}\cos{\beta} e^{a^2 R^2} (r/R)^{2a^2 R^2}},     \\
&(\pa_z r_*)^2=\lambda.
\end{align}
The NHC curves and the closed null curve $g_{\vp\vp}=0$ for both interior and exterior solutions are shown in Fig. \ref{fig:Tipler}.

To conclude, all spacetimes discussed in this subsection have CTCs. Regardless of the positions of the CTCs, i.e., whether they 
are located inside or outside the horizon, the NHCs with minimum separation constant coincide with the $g_{\vp\vp}=0$ closed null curve.

\section{Null Hypersurface Caustics and Super-Entropic Black Holes}\label{3}

Since the super-entropic black hole studied in \cite{2007.04354} is free of closed causal curve, there might be some connection between the presence of NHC outside a black hole horizon and the property that said black hole is super-entropic. This is what we propose to study in this section. Let us start with a brief explanation on what it means to be ``super-entropic''. In the recent years, the inclusion of the negative cosmological constant as a thermodynamical variable (a ``pressure'') in anti-de Sitter spacetime, and its subsequent rich phenomenology, dubbed ``black hole chemistry'', has received a lot of attention (see \cite{kn:chemistry} for a review). Specifically, the thermodynamical pressure is $P=-\Lambda/8\pi G$, where we have restored the Newton's gravitational constant for clarity. The ``thermodynamical volume'' $V$ is defined as the thermodynamic conjugate $\partial M/\partial P$. This notion of volume has no geometric meaning in general, for example, it can even be negative for Taub-NUT \cite{1405.5941}. Nevertheless, for static charged or neutral black holes, it coincides with the naive spherical volume $V=({4}/{3})\pi r_+^3$. There is a conjecture called the ``reverse isoperimetric inequality'' \cite{kn:cvetgib}, which states that the AdS-Schwarzschild black hole has the largest entropy among all black holes with the same thermodynamic volume. In this sense AdS-Schwarzschild is the most stable configuration (under the second law of thermodynamics, the black hole would prefer to evolve towards the state with largest entropy). A ``super-entropic'' black holes are black holes that violate this inequality. There are only a few known super-entropic black holes, so it is possible that the reverse isoperimetric inequality holds for ``most'' black holes (though the exact meaning of ``most'' has yet to be defined; in \cite{1411.4309} it is conjectured that all black holes with compact horizon would satisfy the reverse isoperimetric inequality). 

The super-entropic Kerr-AdS black hole \cite{1411.4309,1504.07529,1401.3107} is constructed by first re-scaling the coordinate $\phi$ into $\tilde{\phi}:=\phi/\Xi$ and then taking the limit $a \to L$ (this is referred to as the ``ultra-spinning'' limit). The new coordinate $\tilde{\phi}$ is then (re-)compactified. Recently there has been some doubt on its true status as a counter-example to the reverse isoperimetric inequality \cite{1911.12817}. Nevertheless, motivated by \cite{2007.04354}, let us examine two other known examples of super-entropic black holes.

\subsection{Ultra-Spinning Kerr-Sen-AdS Black Hole}

Kerr-Sen black hole \cite{sen} is a rotating charged black hole which is an exact solution of the low-energy heterotic string theory (EMDA theory, short for ``Einstein-Maxwell-Dilaton-Axion'' theory), obtained via applying a solution generating technique to the Kerr solution. The Lagrangian of the theory is given as
\beq
{\cal{L}}=\sqrt{-g}\left[R-\df{1}{2}(\pa \phi)^2-e^{-\phi} F^2 -\df{1}{12}e^{-2\phi} H^2\right],
\eeq
where $F^2=F_{\mu\nu}F^{\mu\nu}$ is the square of the Maxwell field tensor, $\phi$ is the dilaton field, and $H:=-e^{2\phi} \star \d\chi$ is a third-rank tensor field. Here $\chi$ denotes the axion pseudoscalar Hodge-dual to $H$.
The solution with the cosmological constant can also be obtained. 

Recently, Wu et al. found that an ultra-spinning Kerr-Sen-AdS$_4$ black hole can be super-entropic but not always so, depending on the values of the parameters \cite{Wu2020}. We shall now check whether super-entropy is related to NHC in this spacetime geometry. 

The metric of the ultraspininng Kerr-Sen AdS$_4$ black hole is 
\begin{flalign}
g[\text{K-S}]=&-\df{\Delta_r}{\Sigma}\left(\d t-L \sin^2{\theta} \d\vp \right)^2+\df{\Sigma}{\Delta_r}\d r^2+\df{\Sigma}{\sin^2{\theta}}\d\theta^2 \\ \notag &+\df{\sin^4{\theta}}{\Sigma}\left[L \d t -(r^2+2br +L^2) \d\vp \right]^2,
\end{flalign}
where $b:=q^2/(2m)$ is the dilatonic scalar charge, while $m$ and $q$ are the mass and electric charge parameter of the black hole, and
\beq
\Delta_r=(r^2+2b r +L^2 )^2 L^{-2}-2mr,\ \ \Sigma=r^2+2b r +L^2 \cos^2{\theta}.
\eeq
{As discussed in \cite{Wu2020}, the black hole is 
super-entropic for $b/L<1$ and not super-entropic (``sub-entropic'') for $b/L >1$. Note that in order to discuss super/sub-entropy, we need to restrict the parameters $m$ and $b$ so that the black hole spacetime still admits a horizon. It can nevertheless be shown that caustics can appear even around naked singularities, but without a horizon we cannot discuss the notion of entropy.}

The $tt$, $t\vp$, and $\vp\vp$ components of the metric are:
\begin{align}
\notag &g_{tt}=\df{1}{\Sigma}\left(-\Delta_r +L^2 \sin^4{\theta} \right),\\
\notag &g_{t\vp}=\df{L \sin^2{\theta}}{\Sigma} \left[\Delta_r -\sin^2{\theta}(r^2+2br + L^2) \right],\\
\notag &g_{\vp \vp}=\df{2mr  L^2}{\Sigma}\sin^4{\theta}.
\end{align}
Since $g_{\vp\vp}$ is always positive, much like the ultraspinning Kerr-AdS black hole, the ultraspinning Kerr-Sen-AdS$_4$ black hole is free of closed causal curve in the entire spacetime.
To calculate $g^{tt}$, we need to compute $D:=g_{t\vp}^2-g_{tt} g_{\vp \vp}$. The expression is simple:
\beq
D=\Delta_r \sin^4{\theta}.
\eeq
Using this, the inverse component $g^{tt}$ is readily obtained to be
\beq
g^{tt}=-\df{g_{\vp\vp}}{D}=-\df{2m r L^2}{\Delta_r \Sigma}.
\eeq
Since the null hypersurface is given by 
\beq
g^{\mu\nu}\pa_\mu (t+r_*) \pa_\nu (t+r_*)=0,
\eeq
we obtain
\beq
-\df{2m r L^2}{\Delta_r}+\Delta_r (\pa_r r_*)^2+\sin^2{\theta} (\pa_\theta r_*)^2=0.
\eeq

\begin{figure}[h!] 
  \centering

  \includegraphics[width=8cm]{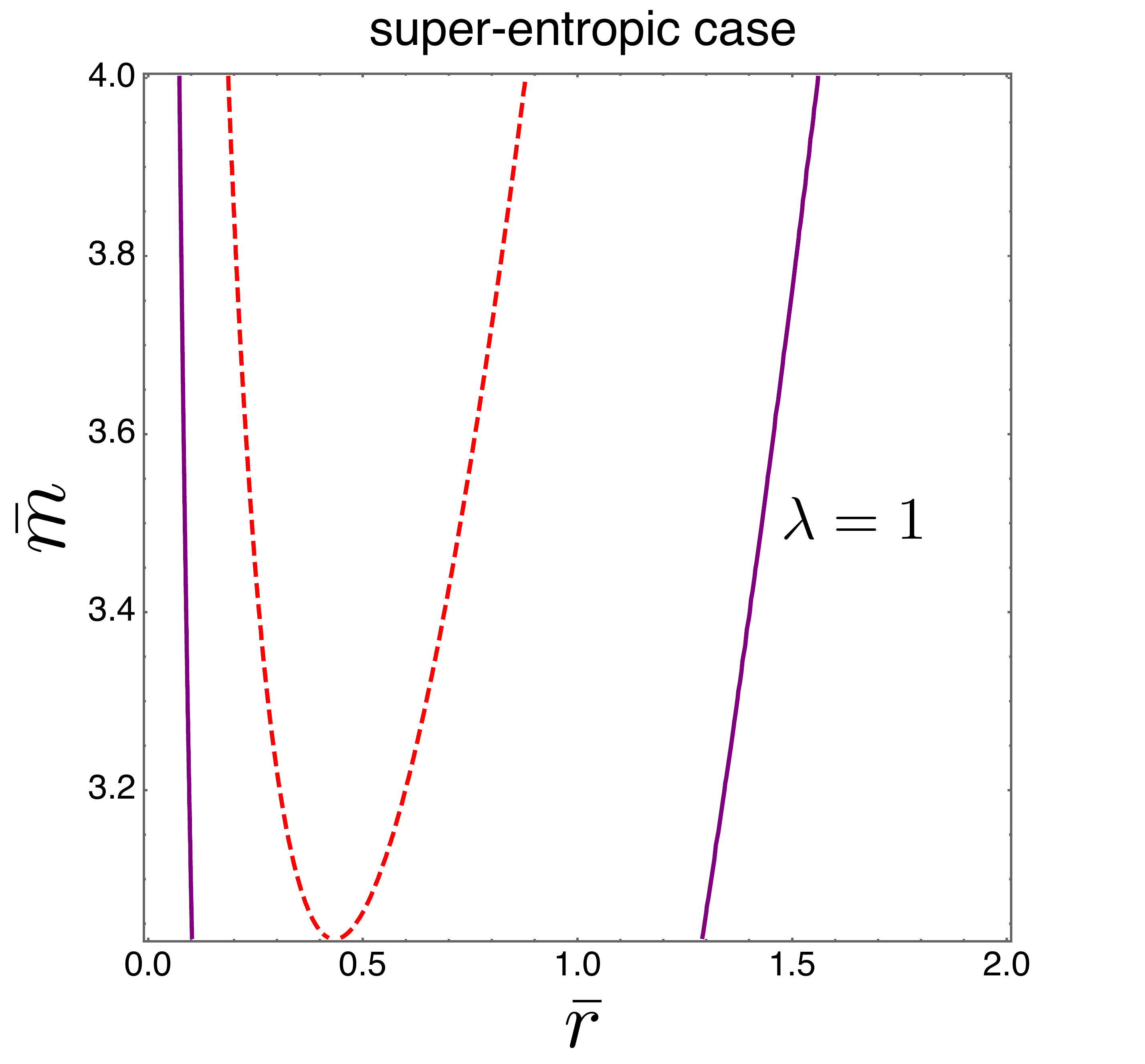}

\bigbreak

  \includegraphics[width=8cm]{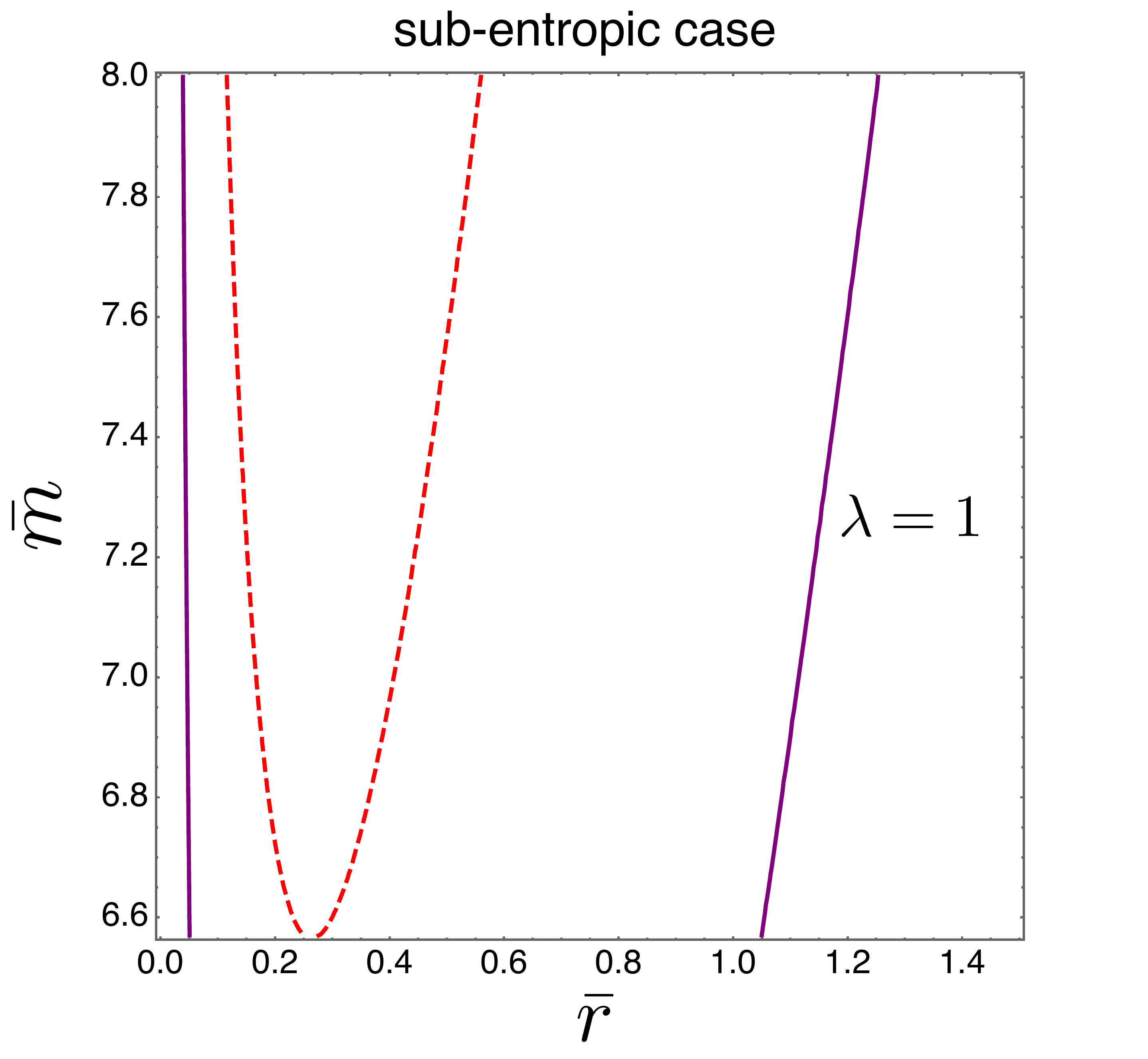}

  \caption{\footnotesize{For the ultra-spinning Kerr-Sen-AdS$_4$ black holes: contours of the caustics $2mr-\lambda \Delta_r=0$ {with $\lambda=1$ (purple)}, and horizon curve (dashed red curve) on the $(\bar{r},\bar{m})$-plane for {super-entropic case ($\bar{b}=0.5$) and sub-entropic case ($\bar{b}=1.5$)}, where quantities with bar are normalized by $L$. {There exist caustics outside the horizon in both super/sub-entropic cases}.}}
  \label{fig:Kerr-Sen}
\end{figure}

As emphasized in \cite{2003.14349}, even in the asymptotically flat case, the Kerr-Sen spacetime is of Petrov Type I \cite{9504139}, thus, \emph{a priori}, one does not expect it to have a Carter-like constant. Nevertheless, the Hamilton-Jacobi equation is separable for the geodesics \cite{1712.06667}.
For our asymptotically AdS case, Introducing a constant for the separation of variables as $L^2 \lambda$, this equation yields
\beq
(\pa_r r_{*})^2=L^2 \df{(2mr  -\lambda \Delta_r)}{\Delta_r^2},\ \ (\pa_\theta r_*)^2=\df{L^2 \lambda}{\sin^2{\theta}}.
\eeq
Therefore the caustics condition is $2mr-\lambda \Delta_r=0$. We plot this condition and the horizon condition $\Delta_r=0$ in Fig.~(\ref{fig:Kerr-Sen}). 

We note that there exist caustics outside the horizon {in both super and sub-entropic cases.} 
{The result seems to indicate that the existence of NHC outside the horizon is not related to whether the black holes is super-entropic or not at least for Kerr-Sen-AdS case. In the following sub-section, we will see that this is also the case for charged BTZ black strings.}



\subsection{Charged BTZ Black String}
Another example of the super-entropic black hole is the charged BTZ spacetime \cite{Johnson2020}. To make a four-dimensional solution\footnote{This is to allow the introduction of the separation constant $\lambda$, for a fair comparison with the other cases we discussed thus far.}, we consider an extra fourth-dimension in the $z$-direction (charged BTZ black string). The metric is written as
\beq
g[\text{BTZ}]=-f(r) \d t^2 +\frac{1}{f(r)}\d r^2 +r^2 \d\vp^2+h(z)\d z^2,
\eeq
where
\beq
f(r)=-8M-\df{{q}^2}{2}\log{(r/L)}+\df{r^2}{L^2},
\eeq
and $h(z)$ is an arbitrary positive definite function. The cosmological constant sets the length scale via $\Lambda:=-1/L^2$. As discussed in the Appendix B, this 
spacetime can be super-entropic, depending on the charge parameter $q$ and the functional form of 
the warp factor $h(z)$. However, the spacetime geometry is always free of closed causal curves since 
$g_{\vp\vp} \geqslant 0$.

\begin{figure}[h!] 
  \centering

  \includegraphics[width=8cm]{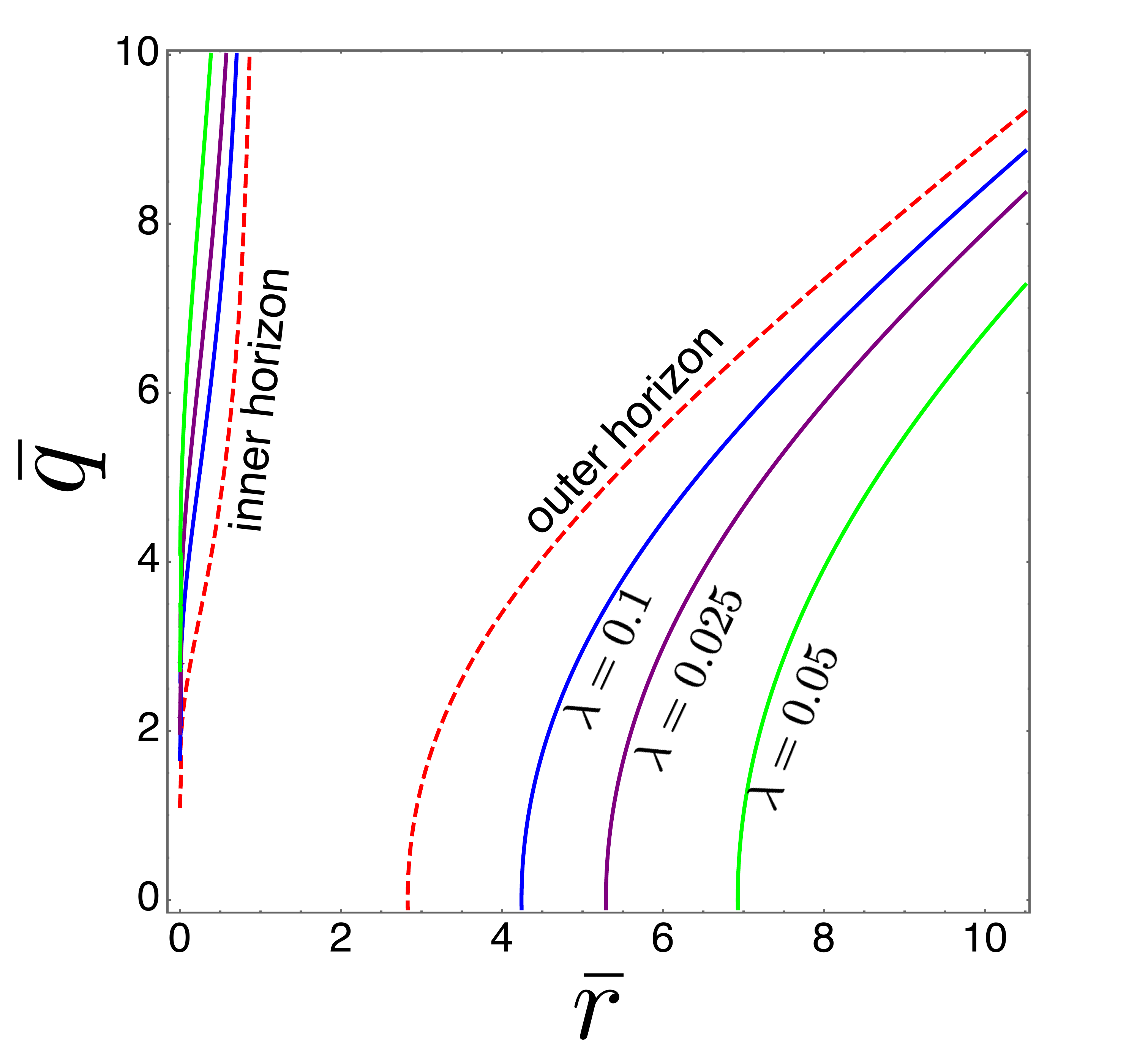}

\bigbreak

  \includegraphics[width=8cm]{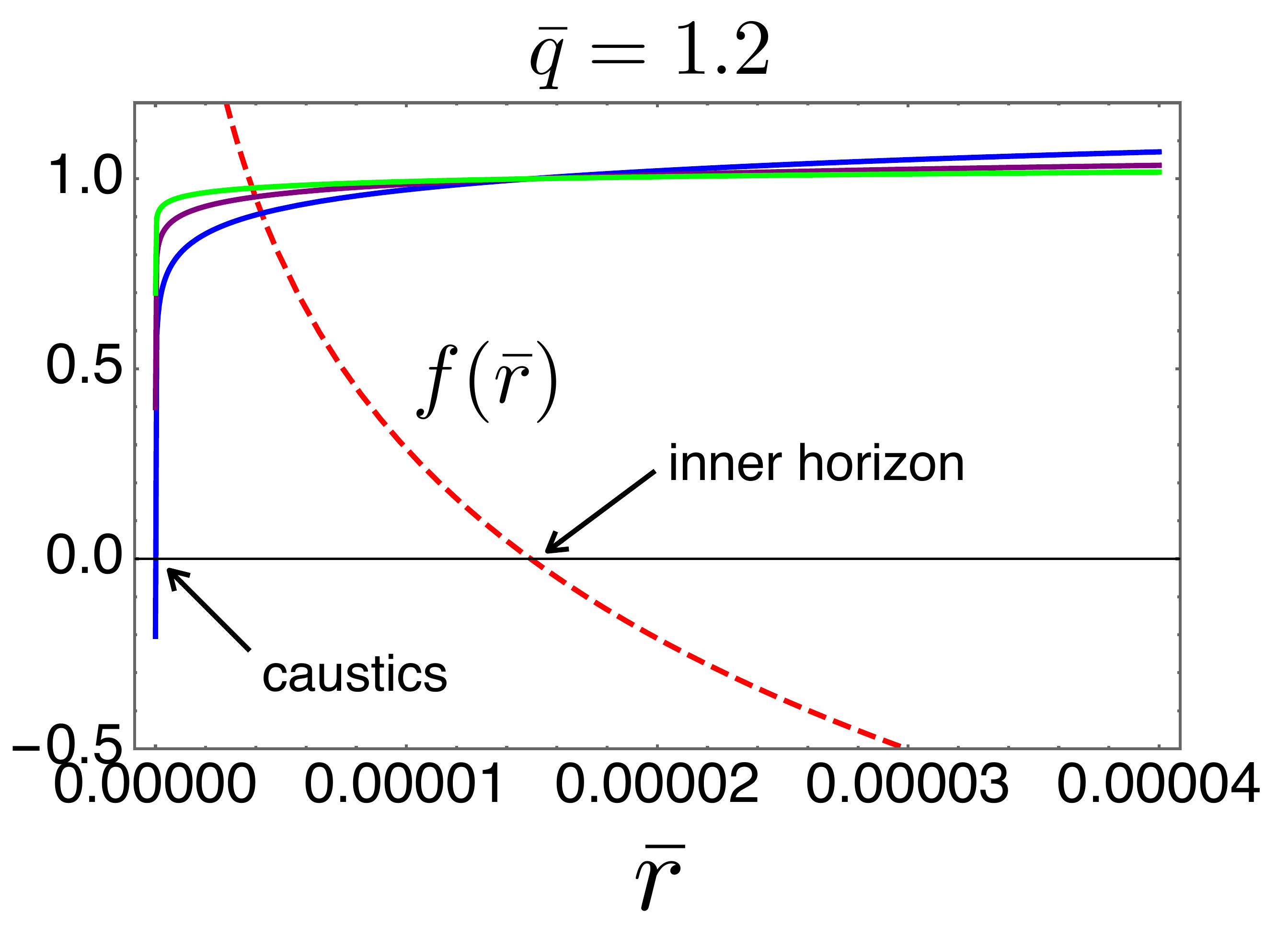}

  \caption{\footnotesize{Contour map of the charged BTZ black string caustics $1-f\lambda=0$ for several values of $\lambda=0.025$ (green), $0.05$ (purple), $0.1$ (blue), and the horizon curve (dashed red curve) on the $(\bar{r},\bar{q})$-plane, where $\bar{r}:=r/L$ and $\bar{q}:=q/L$. 
 The lower panel shows the positions of the inner horizon and caustics for fixed $\bar{q}$ to clarify their behaviors around the inner horizon  (red dashed curve) for small $\bar{q}$ in the upper panel.}}
  \label{fig:rQ}
\end{figure}

To define the null hypersurfaces, we again start by introducing ingoing and outgoing Eddington-Finkelsteing coordinates: 
\beq
\ \ u=t-r_*,\ \ v=t+r_*,
\label{eq:vu}
\eeq
where $r_*=r_* (r,z)$ is the tortoise coordinate. 
The equation that null hypersurface satisfies is 
\beq
g^{\mu\nu}\pa_\mu v \pa_\nu v =g^{tt}+g^{rr}(\pa_r r_{*})^2+g^{zz}(\pa_z r_*)^2=0.
\label{eq:PDE_BTZ}
\eeq
This PDE simplifies into
\beq
-\df{1}{f(r)}+f(r) (\pa_r r_*)^2 +\df{1}{h(z)}(\pa_z r_*)^2=0.
\eeq
Therefore, after introducing the constant for the separation of variables $\lambda$, we can obtain
\beq
(\pa_r r_*)^2=\df{1-f \lambda}{f^2},\ \ (\pa_z r_*)^2=\lambda h(z). 
\eeq
The computation is similar to the other cases explored thus far. The end result yields
\beq
1-f\lambda =0,
\eeq 
which gives the caustics curve. Note that $\lambda\geqslant 0$ since $h(z)>0$.
We plot this condition and the horizon condition $f=0$ on the $(r,q)$-plane in Fig.~\ref{fig:rQ}.

For arbitrary $\lambda$ and arbitrary value of the charge, the caustic curve is outside the horizon. Specifically, and surprisingly, this is also the case for \emph{neutral} BTZ black string. The charged BTZ black string example shows that it is possible that a class of solution can be super-entropic only for some choices of the parameter and metric coefficients, yet there are always NHC present outside the event horizon.

\section{Discussion}\label{4}

In this work, we have discussed the null hypersurface caustics (NHC) of the following spacetimes:
transunital AdS$_5$-Kerr, Taub-NUT, Tipler cylinder, charged (non-rotating) BTZ string, and ultra-spinning Kerr-Sen-AdS$_4$. Among these, transunital black holes, Tipler cylinder, and Taub-NUT spacetimes admit closed timelike curves (CTC), and the NHC with minimal separation constant coincides with the closed null curve, which corresponds to the boundary of the CTC-region. This can occur \emph{outside} the black hole horizon. A general proof of this relationship is difficult. The main idea involved in the calculations is to express $g^{tt}$ as $-g_{\vp\vp}/D$, however the separation of variables requires knowledge on the detailed form of $g_{\vp\vp}$. Such proof should be possible for more restricted class of metric, but then it is doubtful whether such result would shed more light to our understanding.

The relationship between CTC and NHC shows that indeed the caustics in null hypersurfaces reflect the underlying causal structures of the spacetime geometry. In this work we are agnostic about whether CTCs are definitely bad, as it is a matter of ongoing debate, which is outside the scope of our work. (See, e.g., \cite{1912.04702, 1711.08334}, which essentially argued that although point particle can travel on a CTC, any macroscopic objects would be constrained by the second law of thermodynamics. See also \cite{2005.05748}.)

The ultra-spinning Kerr-Sen-AdS$_4$ black holes can be either super-entropic or not, depending on the values of the black hole parameters. {Regardless of the fact that such} black holes are super-entropic, NHCs appear outside their horizon. Together with the ultra-spinning Kerr-AdS$_4$ spacetime, our results suggest that {black holes that can become super-entropic} might {have} NHC outside the horizon. Further evidence is provided by the charged BTZ black string, which is also super-entropic for some choices of the charge parameter and the warp factor. However, for the charged BTZ black string, all of them -- even the sub-entropic ones -- have NHC outside the event horizon. More examples of super-entropic black holes are required to study the relationship between super-entropy and NHC.

\begin{acknowledgments}
YCO thanks the National Natural Science Foundation of China (No.11705162, No.11922508) for funding support. 
SN gratefully acknowledges the hospitality of Kogakuin University, where this work was 
partially done.
\end{acknowledgments}

\begin{appendix}
\section{Derivation of the Metric in Eq. (25)}
Using Eq. \eqref{eq:dr*} and Eq. \eqref{eq:nu}, we obtain
\begin{align}
&\d r=\df{Q \D_r}{Q^2 \D_\theta +P^2 \D_r}\left(\D_\theta \d r_* -\nu P^2 \d\lambda \right),\\
&\d\theta=\df{P \D_\theta}{Q^2 \D_\theta +P^2 \D_r}\left( \D_r \d r_* +\nu Q^2 \d\lambda\right).
\end{align}
Substituting these into $g_{rr}$ and $g_{\theta \theta}$ terms, we get
\begin{align}
\notag &\df{\rho^2}{\D_r} \d r^2 +\df{\rho^2}{\D_\theta} \d\theta^2 \\
&=\df{1}{\Xi^2 R^2}\left( \D_r \D_\theta \d r_*^2 +\nu^2 P^2 Q^2 \d\lambda^2 \right),
\end{align}
whereas the terms with $g_{tt},\ g_{t\vp}$, $g_{\vp\vp}$ can be rewritten as
\begin{align}
\notag &g_{tt}\d t^2 +2g_{t \vp} \d t \d\vp +g_{\vp \vp} \d\vp^2 \\
\notag &=\left(g_{tt}-\df{g_{t \vp}^2}{g_{\vp\vp}}\right)\d t^2
+\df{g_{t \vp}^2}{g_{\vp\vp}}\d t^2+2g_{t \vp} \d t \d\vp +g_{\vp \vp} \d\vp^2\\
&=-\df{D}{g_{\vp\vp}}\d t^2 +g_{\vp\vp}(\d\vp -\omega \d t)^2=-\df{\D_r \D_\theta}{\Xi^2 R^2}\d t^2+g_{\vp\vp}\d\varpi^2.
\end{align}
Thus we obtain Eq. \eqref{eq:metric}.

\section{Is the charged BTZ black string super-entropic?}
Although three-dimensional charged BTZ spacetime is a super-entropic black hole \cite{Johnson2020}, it does not trivially follow that the solution with an ``extra direction'' ($z$-direction) is always 
super-entropic.

Following the procedure which has already been discussed in the literature \cite{kn:cvetgib, 1411.4309}, we evaluate whether the reverse isoperimetric inequality 
\beq
{\cal{R}}:=\left[\frac{(D-1)V}{\omega_{D-2}}\right]^{\frac{1}{D-1}}\left[\frac{\omega_{D-2}}{\cal{A}}\right]^{\frac{1}{D-2}}\geqslant 1,
\eeq
holds. Here $D$ is the spacetime dimension and $\omega_{D}$ stands for the area of the space orthogonal to constant $(t,r)$. 
The horizon area ${\cal{A}}$ can be obtained from the metric as follows
\beq
{\cal{A}}=2\pi r_+ \int_{-\infty}^{+\infty} h(z) dz := 2 \pi r_+ H,
\eeq
where $r_+$ is the radius of the outer horizon. 

One immediately notices that one must constrain the horizon area to be finite in order to have any chance for the black string to be super-entropic.
This means that the warp factor $h(z)$ should be chosen in such a way that the area integral converges. 

Now, $M$ is the energy (mass) of the black hole obtained from the horizon condition $f(r_+)=0$: 
 \beq
 M=\df{r_+^2}{8L^2},
 \label{eq:M}
 \eeq
 and $P$ is the pressure given with the cosmological constant $\Lambda$ as 
 \beq
 P=-\df{\Lambda}{8\pi}=\df{(D-2)(D-1)}{16\pi L^2}.
 \label{eq:P}
 \eeq
 The thermodynamic volume $V$ is computed using the formula in black hole chemistry \cite{kn:chemistry}:
\beq
V=\left(\df{\pa M}{\pa P}\right)_{S,q},
\eeq
 the subscript means that the entropy and charge of the black hole should be fixed when taking the partial derivative. Using \eqref{eq:M} and \eqref{eq:P}, the thermodynamic volume is
\beq
V=-\df{\pi q^2 L^2}{12}+\df{\pi}{3}r_+^2.
\eeq
Therefore, for the present metric ${\cal{R}}$ is obtained as
\beq
{\cal{R}}=\left[ \df{3V}{\omega_2}\right]^{1/3}\left[ \df{\omega_2}{{\cal{A}}} \right]^{1/2}=
H^{-1/3}r_+^{-1/2}\left[\df{r_+^2}{2}- \df{q^2 L^2}{8}\right]^{1/3},
\eeq
where we have $\omega_2=2\pi H$. Depending on the parameter $q$ and the integrated warp factor $H$, the ratio ${\cal{R}}$ can be either larger or smaller than unity. That is, charged BTZ black string can be super-entropic but only for suitable choices of the charge value and the form of the warped function.

\end{appendix}

\end{document}